Title: First principles prediction of the elastic, electronic, and optical properties of Sb2S3 and Sb2Se3 compounds

Corresponding Author: Mr husnu koc, INSTITUTE OF BASIC AND APPLIED SCIENCES

Corresponding Author's Institution: INSTITUTE OF BASIC AND APPLIED SCIENCES

First Author: husnu koc, INSTITUTE OF BASIC AND APPLIED SCIENCES

Order of Authors: husnu koc, INSTITUTE OF BASIC AND APPLIED SCIENCES; Amirullah M mamedov, Prof. Dr.; Engin Deligöz, associate professor

Abstract: We have performed a first principles study of structural, mechanical, electronic, and optical properties of orthorhombic Sb2S3 and Sb2Se3 compounds. The calculations have been carried out within the local density approximation using norm conserving pseudopotentials. The lattice parameters, bulk modulus, and its pressure derivatives of the these compounds have been obtained. The second-order elastic constants have been calculated, and the other related quantities such as the Young's modulus, shear modulus, Poisson's ratio, anisotropy factor, sound velocities and Debye temperature have also been estimated in the present work. The linear photon-energy dependent dielectric functions and some optical properties such as the energy-loss function, the effective number of valance electrons and the effective optical dielectric constant are calculated. Our structural estimation and some other results are in agreement with the available experimental and theoretical data.

we submit our manuscript " First principles prediction of the elastic, electronic, and optical properties of $Sb_2S_3$ and $Sb_2Se_3$ compounds". This manuscript is original and has not been published elsewhere and their intent is to publish in the Solid State Sciences.

Dear Editor,
Thank you for your useful comments and suggestions on the language and structure of our manuscript. We have modified the manuscript accordingly, and detailed corrections are listed below point by point:

1) Please improve the clarity of letters used in figures 1 and 4.

- ✓ the clarity of letters used in figures 1 and 4 have been improved.

2. Graphical abstract should be provided with figure only; the use of text will be allowed only when there is no alternative.

- ✓ Graphical abstract have been provided with figure only.

3.The tel/fax numbers (with country and area code)of the corresponding author should be provided on the first page of the manuscript.

- ✓ On the first page of the manuscript have been provided the tel/fax numbers of the corresponding author.

4. Page should be numbered consecutively.

- ✓ Page have been numbered.

*Graphical Abstract

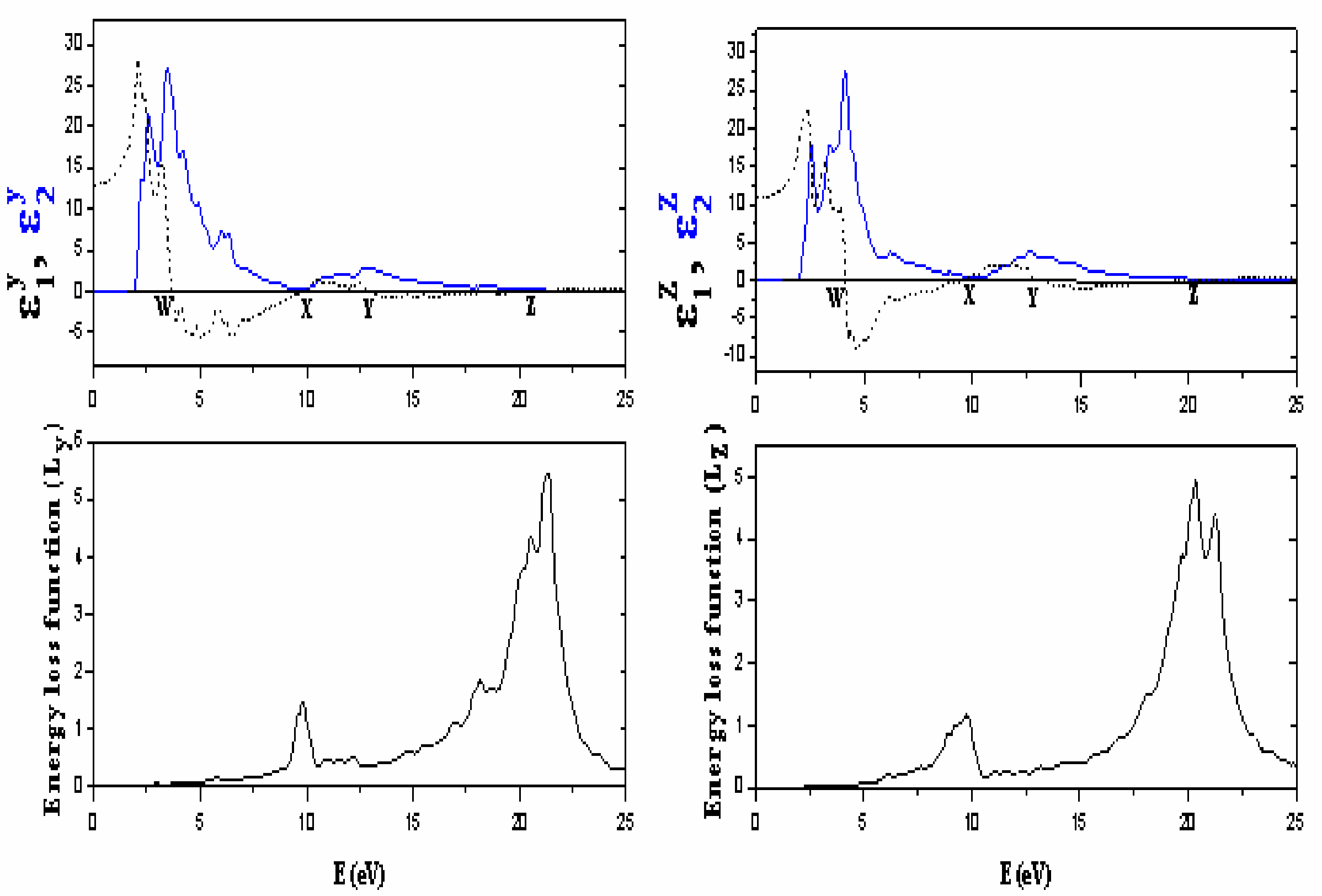

# First principles prediction of the elastic, electronic, and optical properties of $Sb_2S_3$ and $Sb_2Se_3$ compounds

H. Koc[1*], Amirullah M. Mamedov[2], and E. Deligöz[3]

[1]Vocational High School, Siirt University, 56100 Siirt, Turkey
[2]Department of Physics, Cukurova University, Adana, Turkey
[3]Department of Physics, Aksaray University, 68100 Aksaray, Turkey

**Abstract**

We have performed a first principles study of structural, mechanical, electronic, and optical properties of orthorhombic $Sb_2S_3$ and $Sb_2Se_3$ compounds. The calculations have been carried out within the local density approximation using norm conserving pseudopotentials. The lattice parameters, bulk modulus, and its pressure derivatives of the these compounds have been obtained. The second-order elastic constants have been calculated, and the other related quantities such as the Young's modulus, shear modulus, Poisson's ratio, anisotropy factor, sound velocities and Debye temperature have also been estimated in the present work. The linear photon-energy dependent dielectric functions and some optical properties such as the energy-loss function, the effective number of valance electrons and the effective optical dielectric constant are calculated. Our structural estimation and some other results are in agreement with the available experimental and theoretical data.

## 1. Introduction

$Sb_2S_3$ and $Sb_2Se_3$, a member of compounds with the general formula $A_2^V B_3^{VI}$ ($A$=Bi, Sb and $B$=S, Se) are layer structured semiconductors with orthorhombic crystal structure (space group Pnma; No:62), in which each Sb-atom and each Se(S)-atom is bound to three atoms of the opposite kind that are then held together in the crystal by weak secondary bond [1,2]. These crystals have four $Sb_2B_3$ (B=S, Se) molecules (20 atoms) in unit cell. Therefore, these compounds have a complex structure with 56 valance electrons per unit cell. In the last few years, $Sb_2Se_3$ has received a great deal of attention due to its switching effects [3] and its excellent photovoltaic properties and high thermoelectric power [4], which make it possess promising applications in solar selective and decorative coating, optical and thermoelectric cooling devices [5]. On the other hand, $Sb_2S_3$ has attracted attention for its applications as a target material for TV systems [6,7], as well as in microwave [8], switching [9], and optoelectronic devices [10-12].

The positions, used in the computations, for every atom in the unit cell of the orthorhombic crystal system were generated by using the symmetry operation of that system. The position of the $i$th atom ($i=1,2,3,4$) of the atomic species $j$ ($J$ = Sb, S, Bi, Se, I, Cl and Br) in the unit cell of the orthorhombic cry[1]stal system is given by $R_i^{(j)} = R_0 + r_i^{(j)}$, where $R_0$ is the translation vector of that system and represents the considered unit cell. The vector $r_i^{(j)}$ indicates the nonprimitive translation vector, and it is written as $r_i^{(j)} = u_i^{(j)} a_1 + v_i^{(j)} a_2 + \lambda_i^{(j)} a_3$. Here $a_i\ (i = 1,2,3)$ are the primitive translation. The set of the quantities $u_i^{(j)}, v_i^{(j)}$ and $\lambda_i^{(j)}$, the symmetry operations, are given in the following form:

$$(u_i^{(j)}, v_i^{(j)}, \lambda_i^{(j)}) = (x_j, y_j, z_j), (-x_j, -y_j, z_j + 1/2),$$
$$(-x_j + 1/2, y_j + 1/2, z_j + 1/2),$$
$$(x_j + 1/2, -y_j + 1/2, z_j).$$

* Corresponding author. Fax.: +90 2243692.
*E-mail address*: hkoc@student.cu.edu.tr (H. Koc).

The primitive translations and the values of $u_i, v_i$ and $\lambda_i$ are taken in Reference [13-16]

Nasr et al [17] computed the electronic band structure, density of states, charge density and optical properties, such as the dielectric function, reflectivity spectra, refractive index and the loss function using the full potential linearized augmented plane waves (FP-LAPW) method as implemented in the Wien2k code. Kuganathan et al [18] used density functional methods as embedded in the SIESTA code, to test the proposed model theoretically and investigate the perturbations on the molecular and electronic structure of the crystal and the SWNT (single walled carbon nanotubes) and the energy of formation of the $Sb_2Se_3$ SWNT composite. The valance electron density, the electron band structure, and the corresponding electronic density-of-states (DOS) of $A_2B_3$ (A=Bi, Sb and B=S, Se) compounds using the density functional theory were studied by Caracas et al [16].

As far as we know, no ab initio general potential calculations of the elastic constants, Young's modulus, shear modulus, Poisson's ratio, anisotropy factor, sound velocities, Debye temperature, and optical properties such as the energy-loss function, the effective number of valance electrons and the effective optical dielectric constant along y- and z- axes of the $Sb_2S_3$ and $Sb_2Se_3$ have been reported in detail. In the present work, we have investigated the structural, electronic, mechanical, and photon energy-dependent optical properties of the $Sb_2S_3$ and $Sb_2Se_3$ crystals. The method of calculation is given in Section 2; the results are discussed in Section 3. Finally, the summary and conclusion are given in Section 4.

## 2. Method of calculation

The calculations were performed using the density functional formalism and local density approximation (LDA) [19] through the Ceperley and Alder functional [20] as parameterized by Perdew and Zunger [21] for the exchange-correlation energy in the SIESTA code [22,23]. This code calculates the total energies and atomic forces using a linear combination of atomic orbitals as the basis set. The basis set is based on the finite range pseudoatomic orbitals (PAOs) of the Sankey_Niklewsky type [24], generalized to include multiple-zeta decays.

The interactions between electrons and core ions are simulated with separable Troullier-Martins [25] norm-conserving pseudopotentials. We have generated atomic pseudopotentials separately for atoms, Sb, S and Se by using the $5s^2 5p^3$, $3s^2 3p^4$ and $4s^2 4p^4$ configurations, respectively. The cut-off radii for present atomic pseudopotentials are taken as s:1.63 au, p: 1.76 au, 1.94 au for the d and f channels of S, s: 1.94 au, p: 2.14 au, d: 1.94 au. f: 2.49 of Se and 2.35 for the s, p, d and f channels of Sb.

Siesta calculates the self-consistent potential on a grid in real space. The fineness of this grid is determined in terms of an energy cut-off $E_c$ in analogy to the energy cut-off when the basis set involves plane waves. Here by using a double-zeta plus polarization (DZP) orbitals basis and the cut-off energies between 100 and 450 $Ry$ with various basis sets, we found an optimal value of around 350 $Ry$ for $Sb_2S_3$ and $Sb_2Se_3$. For the final computations, 10 k-points for $Sb_2S_3$ and 36 k-points for $Sb_2Se_3$ were enough to obtain the converged total energies ΔE to about 1meV/atoms.

## 3. Results and discussion

### 3.1 Structural properties

All physical properties are related to the total energy. For instance, the equilibrium lattice constant of a crystal is the lattice constant that minimizes the total energy. If the total energy is calculated, any physical property related to the total energy can be determined.

For $Sb_2S_3$ and $Sb_2Se_3$, structures which are orthorhombic are considered. Firstly, the equilibrium lattice parameters, the bulk modulus, and its pressure derivative have been computed minimizing the crystal’s total energy calculated for the different values of lattice constant by means of Murnaghan’s equation of states (eos) [26], and the results are shown in Table 1 along with the experimental and theoretical values. The lattice parameters for $Sb_2S_3$ and $Sb_2Se_3$ are found to be a= 11.311 Å, b=3.834 Å, c=11.224 Å and a=11.71 Å, b=4.14 Å, c=11.62 Å, respectively and they are in a good agreement with the experimental and theoretical values. In all our calculations, we have used the computed lattice parameters. In the present case, the calculated bulk moduli for $Sb_2S_3$ and $Sb_2Se_3$ are 75.730 and 64.78 GPa, respectively. The bulk modulus for $Sb_2S_3$ are higher (about 5.4% ) than the other theoretical result given in Ref [17]. This small difference may stem from the different density- functional- based electronic structure methods.

### 3.2. Elastic properties

The elastic constant of solids provides a link between the mechanical and dynamical behavior of crystals, and give important information concerning the nature of the forces operating in solids. In particular, they provide information on the stability and stiffness of materials, and their ab initio calculation requires precise methods. Since the forces and the elastic constants are functions of the first-order and second-order derivatives of the potentials, their calculation will provide a further check on the accuracy of the calculation of forces in solids. They also provide valuable data for developing inter atomic potentials.

Here, to compute the elastic constants $(C_{ij})$, we have used the "volume-conserving" technique [32]. The present elastic constants for $Sb_2S_3$ and $Sb_2Se_3$ are given in Table 2. Unfortunately, there are no theoretical results for comparing with the present work. Then, our results can serve as a prediction for future investigations.

Nine independent strains are necessary to compute the elastic constants of orthorhombic $Sb_2S_3$ and $Sb_2Se_3$ compounds. Mechanical stability leads to restrictions on the elastic constants, which for orthorhombic crystals [32,33,34] are

$$(C_{11}+C_{22}-2C_{12})>0,\ (C_{11}+C_{33}-2C_{13})>0,$$
$$(C_{22}+C_{33}-2C_{23})>0,\ C_{11}>0, C_{22}>0,$$
$$C_{33}>0, C_{44}>0, C_{55}>0, C_{66}>0, \tag{1}$$
$$(C_{11}+C_{22}+C_{33}+2C_{12}+2C_{13}+2C_{23})>0.$$

The present elastic constants in Table 2 obey these stability conditions for orthorhombic $Sb_2S_3$ and $Sb_2Se_3$. The elastic constants $C_{11}$, $C_{22}$, and $C_{33}$ measure the a-, b-, and c- direction resistance to linear compression, respectively. The calculated $C_{11}$ and $C_{22}$ for $Sb_2S_3$ is lower than the $C_{33}$ while the calculated $C_{33}$ for $Sb_2Se_3$ is lower than the $C_{11}$ and $C_{22}$. Thus, the a axis for $Sb_2S_3$ is more compressible than the c axis while the c axis for $Sb_2Se_3$ is more compressible than the a axis.

It is known that, the elastic constant $C_{44}$ is the most important parameter indirectly governing the indentation hardness of a material. The large $C_{44}$ means a strong ability of resisting the monoclinic shear distortion in (100) plane, and the $C_{66}$ relates to the resistance to shear in the <110> direction. In the present case, $C_{44}$ for $Sb_2S_3$ is higher than $Sb_2Se_3$ compound while $C_{66}$ for $Sb_2Se_3$ is higher than $Sb_2S_3$ compound.

A problem arises when single crystal samples are not available, since it is then not possible to measure the individual elastic constants. Instead, the polycrystalline bulk modulus ($B$) and shear modulus ($G$) may be determined. There are two approximation methods to calculate the polycrystalline modulus, namely, the Voigt method [35] and the Reuss method [36]. For specific cases of orthorhombic lattices, the Reuss shear modulus ($G_R$) and the Voigt shear modulus ($G_V$) are

$$\frac{1}{G_R}=\frac{4}{15}(s_{11}+s_{22}+s_{33})-\frac{4}{15}(s_{12}+s_{13}+s_{23}+\frac{3}{15}(s_{44}+s_{55}+s_{66}) \tag{2}$$

and

$$G_V=\frac{1}{15}(C_{11}+C_{22}+C_{33}-C_{12}-C_{13}-C_{23}+\frac{1}{5}(C_{44}+C_{55}+C_{66}) \tag{3}$$

and the Reuss bulk modulus ($B_R$) and Voight bulk modulus ($B_V$) are defined as

$$B_R = \frac{1}{(s_{11} + s_{22} + s_{33}) + 2(s_{12} + s_{13} + s_{23})} \tag{4}$$

and

$$B_V = \frac{1}{9}(C_{11} + C_{22} + C_{33}) + \frac{2}{9}(C_{12} + C_{13} + C_{23}) \tag{5}$$

In Eq. (2) and (4), the $s_{ij}$ are elastic compliance constants. Using energy considerations Hill [37] proved that the Voigt and Reuss equations represent upper and lower limits of the true polycrystalline constants, and recommended that a practical estimate of the bulk and shear moduli were the arithmetic means of the extremes. Hence, the elastic moduli of the polycrystalline material can be approximated by Hill's average and for shear moduli it is

$$G = \frac{1}{2}(G_R + G_V) \tag{6}$$

and for bulk moduli it is

$$B = \frac{1}{2}(B_R + B_V) \tag{7}$$

The Young's modulus, $E$ , and Poisson's ratio, $v$ , for an isotropic material are given by

$$E = \frac{9BG}{3B + G} \text{ and } v = \frac{3B - 2G}{2(3B + G)} \tag{8}$$

respectively [38,39]. Using the relations given above the calculated bulk modulus, shear modulus, Young's modulus, and Poisson's ratio for $Sb_2S_3$ and $Sb_2Se_3$ are give Table 3.

It is known that isotropic shear modulus and bulk modulus are a measure of the hardness of a solid. The bulk modulus is a measure of resistance to volume change by an applied pressure, whereas the shear modulus is a measure of resistance to reversible deformations upon shear stress [40]. Therefore, isotropic shear modulus is better predictor of hardness than the bulk modulus. The isotropic shear modulus, a measurement of resistance to shape change, is more pertinent to hardness and the larger shear modulus is mainly due to its larger $C_{44}$. The calculated isotropic shear modulus and bulk modulus are 36.43, 64.3 GPa and 33.05, 58.72 GPa for $Sb_2S_3$ and $Sb_2Se_3$, respectively. The values of the bulk moduli indicate that, $Sb_2S_3$ is the least compressible material in the $Sb_2Se_3$ compound. The calculated shear modulus for $Sb_2S_3$ is higher than $Sb_2Se_3$ compound.

According to the criterion in refs. [40,41], a material is brittle (ductile) if the $B/G$ ratio is less (high) than 1.75. The value of the $B/G$ is higher than 1.75 for $Sb_2S_3$ and $Sb_2Se_3$. Hence, these materials behave in a ductile manner.

Young's modulus is defined as the ratio of stress and strain, and used to provide a measure of the stiffness of the solid. The material is stiffer if the value of Young's modulus is high. In this context, due to the higher value of Young's modulus (91.93 GPa) $Sb_2S_3$ compound is relatively stiffer than $Sb_2Se_3$ (83.49 GPa). If the value of E, which has an impact on the ductile, increases, the covalent nature of the material also increases. From Table 3, one can see that E increases as one moves from $Sb_2S_3$ to $Sb_2Se_3$.

The value of the Poisson's ratio is indicative of the degree of directionality of the covalent bonds. The value of the Poisson's ratio is small ( $\upsilon$ =0.1) for covalent materials, whereas for ionic materials a typical value of $\upsilon$ is 0.25 [42]. The calculated Poisson's ratios are about 0.262, 0.263 for $Sb_2S_3$ and $Sb_2Se_3$, respectively. Therefore, the ionic contribution to inter atomic bonding for these compounds is dominant. The $\upsilon$=0.25 and 0.5 are the lower and upper limits, respectively, for central force solids [43]. Our $\upsilon$ values are close to the value of 0.25 indicating inter atomic forces are weightlessly central forces in $Sb_2S_3$ and $Sb_2Se_3$.

Many low symmetry crystals exhibit a high degree of elastic anisotropy [44]. The shear anisotropic factors on different crystallographic planes provide a measure of the degree of anisotropy in atomic bonding in different planes. The shear anisotropic factors are given by

$$A_1 = \frac{4C_{44}}{C_{11} + C_{33} - 2C_{13}} \text{ for the \{100\} plane} \quad (9)$$

$$A_2 = \frac{4C_{55}}{C_{22} + C_{33} - 2C_{23}} \text{ for the \{010\}plane} \quad (10)$$

$$A_3 = \frac{4C_{66}}{C_{11} + C_{22} - 2C_{12}} \text{ for the \{001\}plane} \quad (11)$$

The calculated $A_1, A_2$ and $A_3$ for $Sb_2S_3$ and $Sb_2Se_3$ are given in Table 4. A value of unity means that the crystal exhibits isotropic properties while values other than unity represent varying degrees of anisotropy. From Table 4, it can be seen that $Sb_2S_3$ and $Sb_2Se_3$ exhibit low anisotropy. Another way of measuring the elastic anisotropy is given by the percentage of anisotropy in the compression and shear [38, 39, 45].

$$A_{comp} = \frac{B_V - B_R}{B_V + B_R} x100 \quad (12)$$

$$A_{shear} = \frac{G_V - G_R}{G_V + G_R} x100 \quad (13)$$

For crystals, these values can range from zero (isotropic) to 100% representing the maximum anisotropy. The percentage anisotropy values have been computed for $Sb_2S_3$ and $Sb_2Se_3$, and are shown in Table 4. For these compounds, it can be seen that the anisotropy in compression is small and the anisotropy in shear is high. $Sb_2S_3$ compound exhibit relatively high shear and bulk anisotropies among these compounds.

The Debye temperature is known as an important fundamental parameter closely related to many physical properties such as specific heat and melting temperature. At low temperatures the vibrational excitations arise solely from acoustic vibrations. Hence, at low temperatures the Debye temperature calculated from elastic constants is the same as that determined from specific heat measurements. We have calculated the Debye temperature, $\theta_D$, from the elastic constants data using the average sound velocity, $v_m$, by the following common relation given in Ref. [46]

$$\theta_D = \frac{\hbar}{k}\left[\frac{3n}{4\pi}\left(\frac{N_A\rho}{M}\right)\right]^{1/3} v_m, \tag{14}$$

where $\hbar$ is Planck's constants, $k$ is Boltzmann's constant, $N_A$ is Avogadro's number, n is the number of atoms per formula unit, $M$ is the molecular mass per formula unit, $\rho(=M/V)$ is the density, and $v_m$ is given [47] as

$$v_m = \left[\frac{1}{3}\left(\frac{2}{v_t^3}+\frac{1}{v_l^3}\right)\right]^{-1/3}, \tag{15}$$

where $v_l$ and $v_t$, are the longitudinal and transverse elastic wave velocities, respectively, which are obtained from Navier's equation [48]

$$v_l = \sqrt{\frac{3B+4G}{3\rho}}, \tag{16}$$

and

$$v_t = \sqrt{\frac{G}{\rho}} \tag{17}$$

The calculated values of the longitudinal, transverse, average sound velocities and density in the present formalism are shown in Table 5 along with the Debye temperature. The calculated Debye temperature for $Sb_2S_3$ is higher than $Sb_2Se_3$. Unfortunately, there are no theoretical and experimental results to compare with the calculated $v_l$, $v_t$, $v_m$ , and $\theta_D$ values.

### 3.3. Electronic properties

For a better understanding of the electronic and optical properties of $Sb_2S_3$ and $Sb_2Se_3$, the investigation of the electronic band structure would be useful. The electronic band structures of orthorhombic $Sb_2S_3$ and $Sb_2Se_3$ single crystals have been calculated along high symmetry directions in the first Brillouin zone (BZ). The band structures were calculated along the special lines connecting the high-symmetry points S (1/2,1/2,0), Y (0,1/2,0), $\Gamma$ (0,0,0), S(1/2,1/2,0), R (1/2,1/2,1/2) for $Sb_2S_3$ and $Sb_2Se_3$ in the k-space. The results of the calculation are shown in Fig. 1 for these single crystals.

The energy band structures calculated using LDA for $Sb_2S_3$ and $Sb_2Se_3$ are shown in Fig. 1. As can be seen in Fig. 1a, the $Sb_2S_3$ compound has an direct band gap semiconductor with the value 0.98 $eV$. The top of the valance band and the bottom of the conduction band positioned at the $\Gamma$ point of BZ. The estimates of the band gap of $Sb_2S_3$ are contradictory in the literature. The band gap values estimated for $Sb_2S_3$ vary from 1.56 $eV$ to 2.25 $eV$ (see Table 6). In conclusion, our band gap value obtained is different from experimental and theoretical values and the band gap has same character with given in Ref. [51-54,17]. The present band and the density of states (DOS) profiles for $Sb_2S_3$ agree with the earlier work [17]

The calculated band structure of $Sb_2Se_3$ is given in Fig. 1b. As can be seen from the figure., the band gap has the different character with that of $Sb_2S_3$, that is, it is an indirect band gap semiconductor. The top of the valance band positioned at the nearly midway between $\Gamma$ and S point of BZ, the bottom of the conduction band is located at the nearly midway between the $\Gamma$ and Y point of BZ. The indirect and direct band gap values of $Sb_2Se_3$ compound are, 0.99 $eV$ and 1.07 $eV$, respectively. The band gap values estimated for $Sb_2Se_3$ vary from 1.56 $eV$ to 2.25 $eV$ (see Table 6). Our band gap value obtained is good agreement with experimental and theoretical values and the character of the band gap is different from that given in Ref. [55,56].

In the rightmost panels of this figure, the density of states (DOS) are presented for each crystals. In this figure, the lowest valence bands occur between about -14 and -12 $eV$ are dominated by S 3s and Se 4s states while valence bands occur between about -10 and -7 $eV$ are dominated by Sb 5s states. The other valance bands are essentially dominated by S 3p and Se 4p states. The 5p states of Sb atoms are also contributing to the valance bands, but the values of densities of these states are small compared to S 3p and Se 4p states. The energy region just above Fermi energy level is dominated by Sb 5p.

Band structures of $Sb_2S_3$ and $Sb_2Se_3$ single crystals are compared, band structures of these crystals are highly resemble one another. Thus, on formation of the band structures of $Sb_2S_3$ and $Sb_2Se_3$ the 5s 5p orbitals of Sb atoms are more dominant than 3s3p and 4s4p orbitals of S and Se atoms. Finally, the band gap values obtained are less than the estimated experimental and theoretical results. For all crystal structures considered, the band gap values are underestimated than the experimental values. This state is caused from the exchange-correlation approximation of DFT.

## 3.4. Optical properties

It is well known that the effect of the electric field vector, $\mathbf{E}(\omega)$, of the incoming light is to polarize the material. At the level of linear response, this polarization can be calculated using the following relation [58]:

$$P^i(\omega)=\chi_{ij}^{(1)}(-\omega,\omega)E^j(\omega), \tag{18}$$

where $\chi_{ij}^{(1)}$ is the linear optical susceptibility tensor and it is given by [59]

$$\chi_{ij}^{(1)}(-\omega,\omega)=\frac{e^2}{\hbar\Omega}\sum_{nm\vec{k}} f_{nm}(\vec{k})\frac{r_{nm}^i(\vec{k})r_{mn}^i(\vec{k})}{\omega_{mn}(\vec{k})-\omega}=\frac{\varepsilon_{ij}(\omega)-\delta_{ij}}{4\pi} \tag{19}$$

where $n,m$ denote energy bands, $f_{mn}(\vec{k})\equiv f_m(\vec{k})-f_n(\vec{k})$ is the Fermi occupation factor, $\Omega$ is the normalization volume. $\omega_{mn}(\vec{k})\equiv\omega_m(\vec{k})-\omega(\vec{k})$ are the frequency differences, $\hbar\omega_n(\vec{k})$ is the energy of band $n$ at wave vector **k.** The $\vec{r}_{nm}$ are the matrix elements of the position operator [59].

As can be seen from Eq. (19), the dielectric function $\varepsilon_{ij}(\omega)=1+4\pi\chi_{ij}^{(1)}(-\omega,\omega)$ and the imaginary part of $\varepsilon_{ij}(\omega)$, $\varepsilon_2^{ij}(\omega)$, is given by

$$\varepsilon_2^{ij}(w)=\frac{e^2}{\hbar\pi}\sum_{nm}\int d\vec{k}f_{nm}(\vec{k})\frac{v_{nm}^i(\vec{k})v_{nm}^j(\vec{k})}{\omega_{mn}^2}\delta(\omega-\omega_{mn}(\vec{k})). \tag{20}$$

The real part of $\varepsilon_{ij}(\omega)$,$\varepsilon_1^{ij}(\omega)$, can be obtained by using the Kramers-Kroning transformation [59]. Because the Kohn-Sham equations determine the ground state properties, the unoccupied conduction bands as calculated have no physical significance. If they are used as single-particle states in a calculation of optical properties for semiconductors, a band gap problem comes into included in calculations of response. In order to take into account self-energy effects, in the present work, we used the 'scissors approximation' [58].

The known sum rules [60] can be used to determine some quantitative parameters, particularly the effective number of the valence electrons per unit cell $N_{eff}$, as well as the effective optical dielectric constant $\varepsilon_{eff}$, which make a contribution to the optical constants of a crystal at the energy $E_0$. One can obtain an estimate of the distribution of oscillator strengths for both intraband and interband transitions by computing the $N_{eff}(E_0)$ defined according to

$$N_{eff}(E)=\frac{2m\varepsilon_0}{\pi\hbar^2e^2Na}\int_0^\infty\varepsilon_2(E)EdE, \tag{21}$$

Where $N_a$ is the density of atoms in a crystal, $e$ and $m$ are the charge and mass of the electron, respectively and $N_{eff}(E_0)$ is the effective number of electrons contributing to optical transitions below an energy of $E_0$.

Further information on the role of the core and semi-core bands may be obtained by computing the contribution which the various bands make to the static dielectric constant, $\varepsilon_0$. According to the Kramers-Kronig relations, one has

$$\varepsilon_0(E)-1=\frac{2}{\pi}\int_0^\infty\varepsilon_2(E)E^{-1}dE. \tag{22}$$

One can therefore define an 'effective' dielectric constant, which represents a different mean of the interband transitions from that represented by the sum rule, Eq. (22), according to the relation

$$\varepsilon_{eff}(E)-1=\frac{2}{\pi}\int_0^{E_0}\varepsilon_2(E)E^{-1}dE. \tag{23}$$

The physical meaning of $\varepsilon_{eff}$ is quite clear: $\varepsilon_{eff}$ is the effective optical dielectric constant governed by the interband transitions in the energy range from zero to $E_0$, i.e. by the polarizition of the electron shells.

In order to calculate the optical response by using the calculated band structure, we have chosen a photon-energy range of 0-25 $eV$ and have seen that a 0-17 $eV$ photon-energy range is sufficient for most optical functions.

The $Sb_2S_3$ and $Sb_2Se_3$ single crystals have an orthorhombic structure that is optically a biaxial system. For this reason, the linear dielectric tensor of the $Sb_2S_3$ and $Sb_2Se_3$ compounds have three independent components that are the diagonal elements of the linear dielectric tensor.

We first calculated the real and imaginary parts of the y- and z-components of the frequency-dependent linear dielectric function and these are shown in Fig. 2 and Fig. 3. The $\varepsilon_1^y$ is equal to zero at about 3.67 $eV$, 9.78 $eV$, 13.02 $eV$, and 20.4 $eV$, ( at the W, X, Y, and Z points in Fig. 2), whereas the other function $\varepsilon_1^z$ is equal to zero at about 4.12 $eV$, 9.61 $eV$, 13.37 $eV$ and 20.1 $eV$ (at the W, X, Y, Z points in Fig. 2) for $Sb_2S_3$ compound. The $\varepsilon_1^y$ is equal to zero at about 2.76 $eV$, 8.77 $eV$, 12.63 $eV$, and 20.04 $eV$, ( at the W, X, Y, and Z points in Fig. 3), whereas the other function $\varepsilon_1^z$ is equal to zero at about 2.95 $eV$, 8.91 $eV$, 13.06 $eV$ and 19.89 $eV$ (at the W, X, Y, Z points in Fig. 3) for $Sb_2Se_3$ compound. The peaks of the $\varepsilon_2^y$ and $\varepsilon_2^z$ correspond to the optical transitions from the valence band to the conduction band and are in agreement with the previous results. The maximum peak values of $\varepsilon_2^y$ and $\varepsilon_2^z$ for $Sb_2S_3$ are around 3.46 $eV$ and 4.1 $eV$, respectively, whereas the maximum values of $\varepsilon_2^y$ and $\varepsilon_2^z$ for $Sb_2Se_3$ are around 2.74 $eV$ and 2.78 $eV$, respectively. Spectral dependences of dielectric functions show the similar features for both materials because the electronic configurations of Se ([Ar],$3d^{10}$ $4s^2$ $4p^2$) and S([Ne], $3s^2$ $3p^3$) are very close to each other. In general, there are various contributions to the dielectric function, but Fig. 2 and Fig. 3 show only the contribution of the electronic polarizability to the dielectric function. The maximum peak values of $\varepsilon_2^y$ and $\varepsilon_2^z$ are in agreement with maximum peak values of theoretical for $Sb_2S_3$ [17]. In the range between 2 $eV$ and 5 $eV$, $\varepsilon_1^z$ decrease with increasing photon-energy, which is characteristics of an anomalous dispersion. In this energy range, the transitions between occupied and unoccupied states mainly occur between S 3p and Se 4p states which can be seen in the DOS displayed in Fig. 1. Furthermore as can be seen from Fig. 2 and Fig. 3, the photon –energy range up to 1.5 $eV$ is characterized by high transparency, no absorption and a small reflectivity. The 1.9-5.0 $eV$ photon energy range is characterized by strong absorption and appreciable reflectivity.

The calculated energy-loss functions, $L(\omega)$, are also presented in Fig. 2 and Fig. 3. In this figure, $L_y$ and $L_z$ correspond to the energy-loss functions along the y- and z-directions. The function $L(\omega)$ describes the energy loss of fast electrons traversing the material. The sharp maxima in the energy-loss function are associated with the existence of plasma oscillations [61]. The curves of $L_y$ and $L_z$ in Fig. 2 and Fig. 3 have a maximum near 21.28 and 20.30 $eV$ for $Sb_2S_3$, respectively and 21.90 and 20.16 $eV$ for $Sb_2Se_3$, respectively. These values coincide with the Z point in figure 2 and figure 3. The maximum piks of energy-loss functions are in agreement with maximum peaks of theoretical for $Sb_2S_3$.

The calculated effective number of valence electrons $N_{eff}$ and the effective dielectric constant $\varepsilon_{eff}$ are given in Fig. 4. The effective number of valence electron per unit cell, $N_{eff}$, contributing in the interband transitions, reaches saturation value at about 20 $eV$. This means that deep-lying valence orbitals paticipate in the interband transitions as well (see Figure 4)

The effective optical dielectric constant, $\varepsilon_{eff}$, shown in Fig. 4, reaches a saturation value at about 20 $eV$. The photon-energy dependence of $\varepsilon_{eff}$ can be separated into two regions. The first is characterized by a rapid rise and it extends up to 12 $eV$. In the second region the value of $\varepsilon_{eff}$ rises more smoothly and slowly and tends to saturation at the energy 20 $eV$. This means that the greatest conribution to $\varepsilon_{eff}$ arises from interband transitions between 2 $eV$ and 17 $eV$.

## Conclusion

In present work, we have made a detailed investigation of the structural, electronic, mechanical, and frequency-dependent linear optical properties of the $Sb_2S_3$ and $Sb_2Se_3$ crystals using the density functional methods. The results of the structural optimization implemented using the LDA are in good agreement with the experimental results. From the present results, we observe that these compounds in mechanically stable structures, The mechanical properties like shear modulus, Young's modulus, Poisson's ratio, Debye temperature, and shear anisotropic factors are also calculated. $Sb_2S_3$ and $Sb_2Se_3$ compounds will behave in a ductile manner. Moreover, the ionic contribution to inter atomic bonding for these compounds is dominant. We have revealed that the orthorhombic $Sb_2S_3$ and $Sb_2Se_3$ compounds are in the ground-state configuration and the band structures of these compounds are semiconductor in nature. We have examined photon-energy dependent dielectric functions, some optical properties such as the energy-loss function, the effective number of valance electrons and the effective optical dielectric constant along the y- and z- axes, and mechanical properties. Since there are no experimental elastic data available for $Sb_2S_3$ and $Sb_2Se_3$ compound, we think that the ab initio theoretical estimation is the only reasonable tool for obtaining such important information.

## Tables with captions

**Table 1.** The calculated equilibrium lattice parameters (a, b, and c), bulk modulus ( $B$ ), and the pressure derivative of bulk modulus ( $B'$ ) together with the theoretical and experimental values for $Sb_2S_3$ and $Sb_2Se_3$ in fractional coordinate.

| Material | Reference | $a$ (Å) | $b$ (Å) | $c$ (Å) | $B(GPa)$ | $B'$ | Space Group |
|---|---|---|---|---|---|---|---|
| $Sb_2S_3$ | Present (LDA) | 11.311 | 3.834 | 11.224 | 75.730 | 4.38 | Pnma (No:62) |
| | Theory (EV-GGA) [a] | 11.304 | 3.836 | 11.216 | 71.624 | 5.00 | |
| | Experimental[b] | 11.311 | 3.839 | 11.223 | | | |
| | Experimental[c] | 11.30 | 3.83 | 11.22 | | | |
| | Experimental[d] | 11.27 | 3.84 | 11.29 | | | |
| $Sb_2Se_3$ | Present (LDA) | 11.71 | 4.14 | 11.62 | 64.78 | 4.75 | Pnma (No:62) |
| | Theory (GGA)[e] | 11.91 | 3.98 | 11.70 | | | |
| | Experimental[f] | 11.79 | 3.98 | 11.64 | | | |
| | Experimental[g] | 11.78 | 3.99 | 11.63 | | | |
| | Experimental[h] | 11.77 | 3.96 | 11.62 | | | |

[a]Reference [17]

[b]Reference [27]

[c]Reference [28]

[d]Reference [29]

[e]Reference [18]

[f]Reference [2]

[g]Reference [30]

[h]Reference [31]

**Table 2.** The calculated elastic constants (in GPa) for $Sb_2S_3$ and $Sb_2Se_3$

| Material | Reference | $C_{11}$ | $C_{22}$ | $C_{33}$ | $C_{12}$ | $C_{13}$ | $C_{23}$ | $C_{44}$ | $C_{55}$ | $C_{66}$ |
|---|---|---|---|---|---|---|---|---|---|---|
| $Sb_2S_3$ | Present (LDA) | 97.5 | 111.78 | 110.72 | 20.81 | 57.94 | 81.83 | 69.11 | 48.0 | 26.68 |
| $Sb_2Se_3$ | Present (LDA) | 101.56 | 89.94 | 84.60 | 34.13 | 43.66 | 48.58 | 54.92 | 40.84 | 30.37 |

**Table 3.** The calculated isotropic bulk modulus (B, in GPa), shear modulus (G, in GPa), Young's modulus (E, in GPa) and Poisson's ratio for $Sb_2S_3$ and $Sb_2Se_3$ compounds.

| Material | Reference | $B_R$ | $B_V$ | $B$ | $G_R$ | $G_V$ | $G$ | $E$ | $\upsilon$ |
|---|---|---|---|---|---|---|---|---|---|
| $Sb_2S_3$ | Present (LDA) | 62.86 | 65.74 | 64.3 | 32.35 | 40.50 | 36.43 | 91.93 | 0.262 |
| $Sb_2Se_3$ | Present (LDA) | 58.72 | 58.76 | 58.74 | 30.89 | 35.21 | 33.05 | 83.49 | 0.263 |

**Table 4.** The shear anisotropic factors $A_1$, $A_2$, $A_3$, and $A_{comp(\%)}$, $A_{shear(\%)}$.

| Material | Reference | $A_1$ | $A_2$ | $A_3$ | $A_{comp}(\%)$ | $A_{shear}(\%)$ |
|---|---|---|---|---|---|---|
| $Sb_2S_3$ | Present (LDA) | 2.41 | 2.37 | 0.68 | 2.23 | 11.2 |
| $Sb_2Se_3$ | Present (LDA) | 2.22 | 2.11 | 0.99 | 0.03 | 6.5 |

**Table 5.** The density, longitudinal, transverse, and average elastic wave velocities, together with the Debye temperature for $Sb_2S_3$ and $Sb_2Se_3$.

| Material | Referance | $\rho(g/cm^3)$ | $v_l(m/s)$ | $v_t(m/s)$ | $v_m(m/s)$ | $\theta_D(K)$ |
|---|---|---|---|---|---|---|
| $Sb_2S_3$ | Present (LDA) | 4.64 | 4932.15 | 2802.02 | 3115.08 | 320.14 |
| | Experimental[b] | 4.61 | | | | |
| $Sb_2Se_3$ | Present (LDA) | 5.67 | 4258.13 | 2414.32 | 2684.51 | 262.78 |
| | Experimental[b] | 5.88 | | | | |

**Table 6.** Energy band gap for $Sb_2S_3$ and $Sb_2Se_3$

| Material | | $E_g\,(eV)$ |
|---|---|---|
| $Sb_2S_3$ | Present | 0.98 direct |
| $Sb_2S_3$ | Experimental[a] | 1.78 indirect-2.25 direct |
| $Sb_2S_3$ | Experimental[b] | 1.63 indirect-1.72 direct |
| $Sb_2S_3$ | Experimental[c] | 1.56 direct |
| $Sb_2S_3$ | Experimental[d] | 1.64 |
| $Sb_2S_3$ | Experimental[e] | 1.71 |
| $Sb_2S_3$ | Experimental[f] | 2.2 (300 K)-1.60 (473 K) direct |
| $Sb_2S_3$ | Theory[g] | 1.55 |
| $Sb_2S_3$ | Theory[h] | 1.76 |
| | | |
| $Sb_2Se_3$ | Present | 0.99 indirect-1.07 direct |
| $Sb_2Se_3$ | Experimental[ı] | 1.0 direct |
| $Sb_2Se_3$ | Experimental[i] | 1.5 direct |
| $Sb_2Se_3$ | Experimental[j] | 1.1 indirect |
| $Sb_2Se_3$ | Experimental[k] | 1.0-1.2 indirect |
| $Sb_2Se_3$ | Theory[g] | 1.14 |

[a]Reference [49]

[b]Reference [50]

[c]Reference [51]

[d]Reference [52]

[e]Reference [53]

[f]Reference [54]

[g]Reference [16]

[h]Reference [17]

[ı]Reference [55]

[i]Reference [56]

[j]Reference [51]

[k]Reference [57]

Figure

## Figure captions

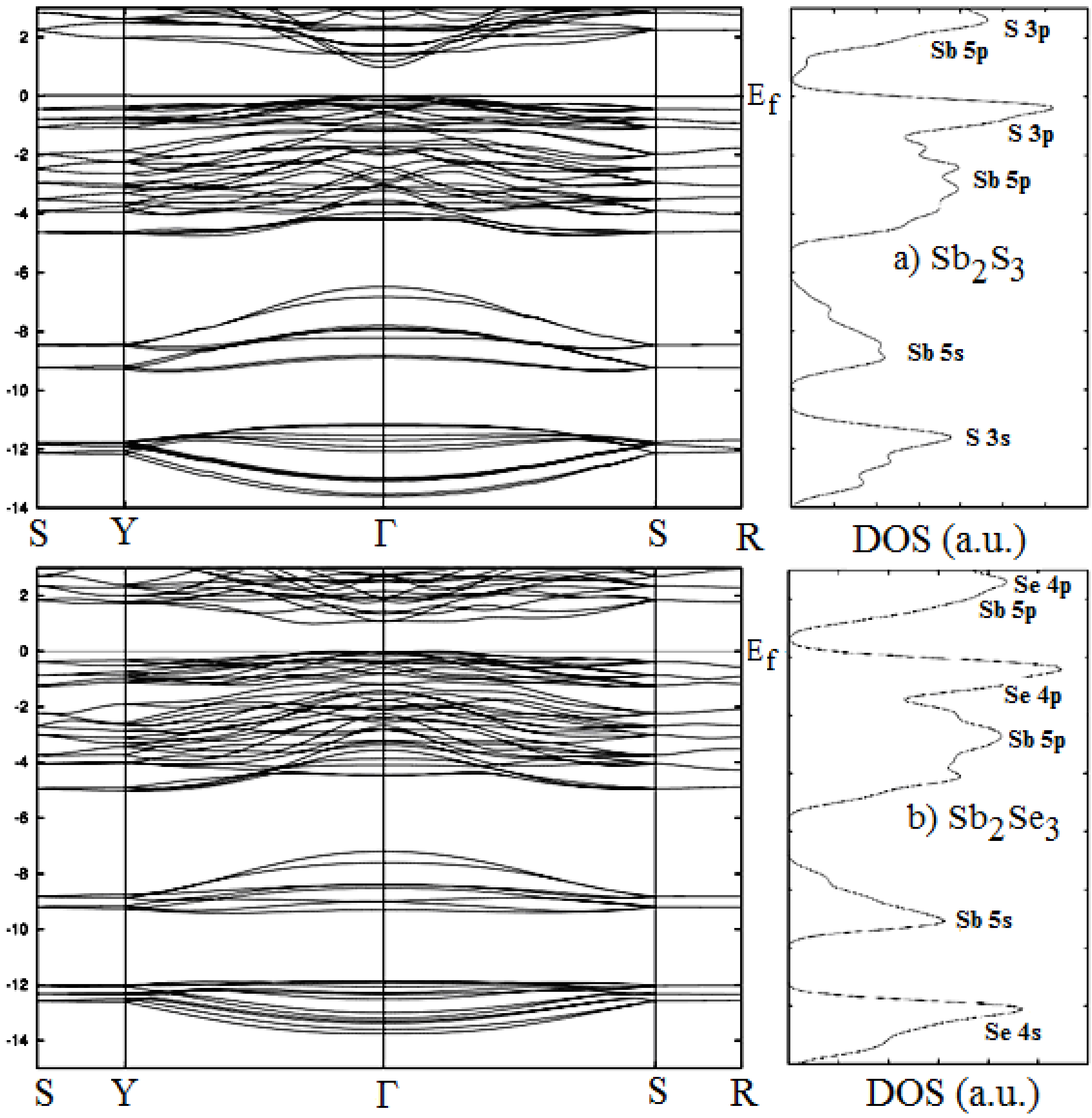

**Fig 1.** Energy band structure and DOS ( density of states) for $Sb_2S_3$ and $Sb_2Se_3$.

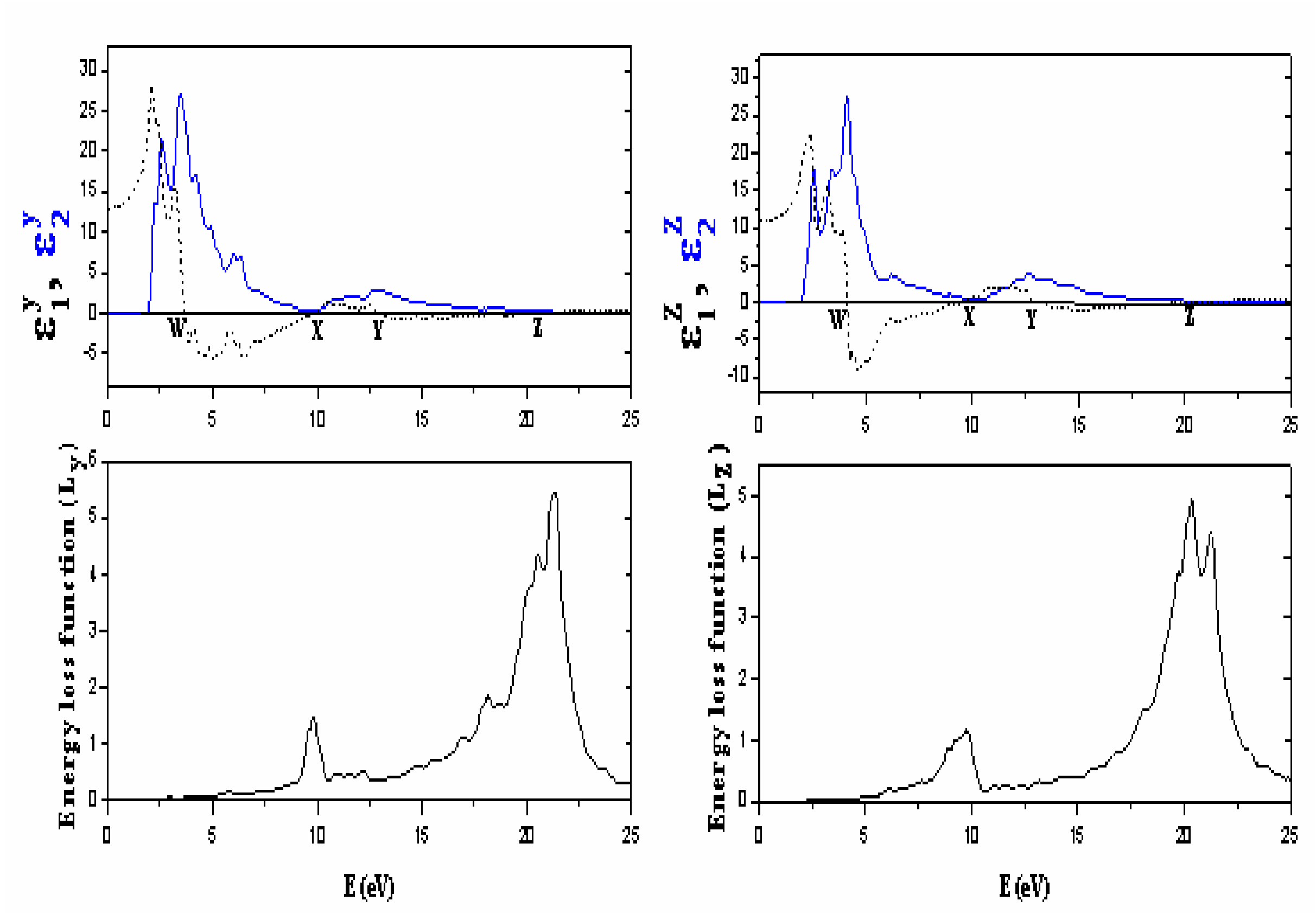

**Fig 2**. Energy spectra of dielectric function $\varepsilon = \varepsilon_1 - i\varepsilon_2$ and energy-loss function (L) along the y- and z-axes for $Sb_2S_3$

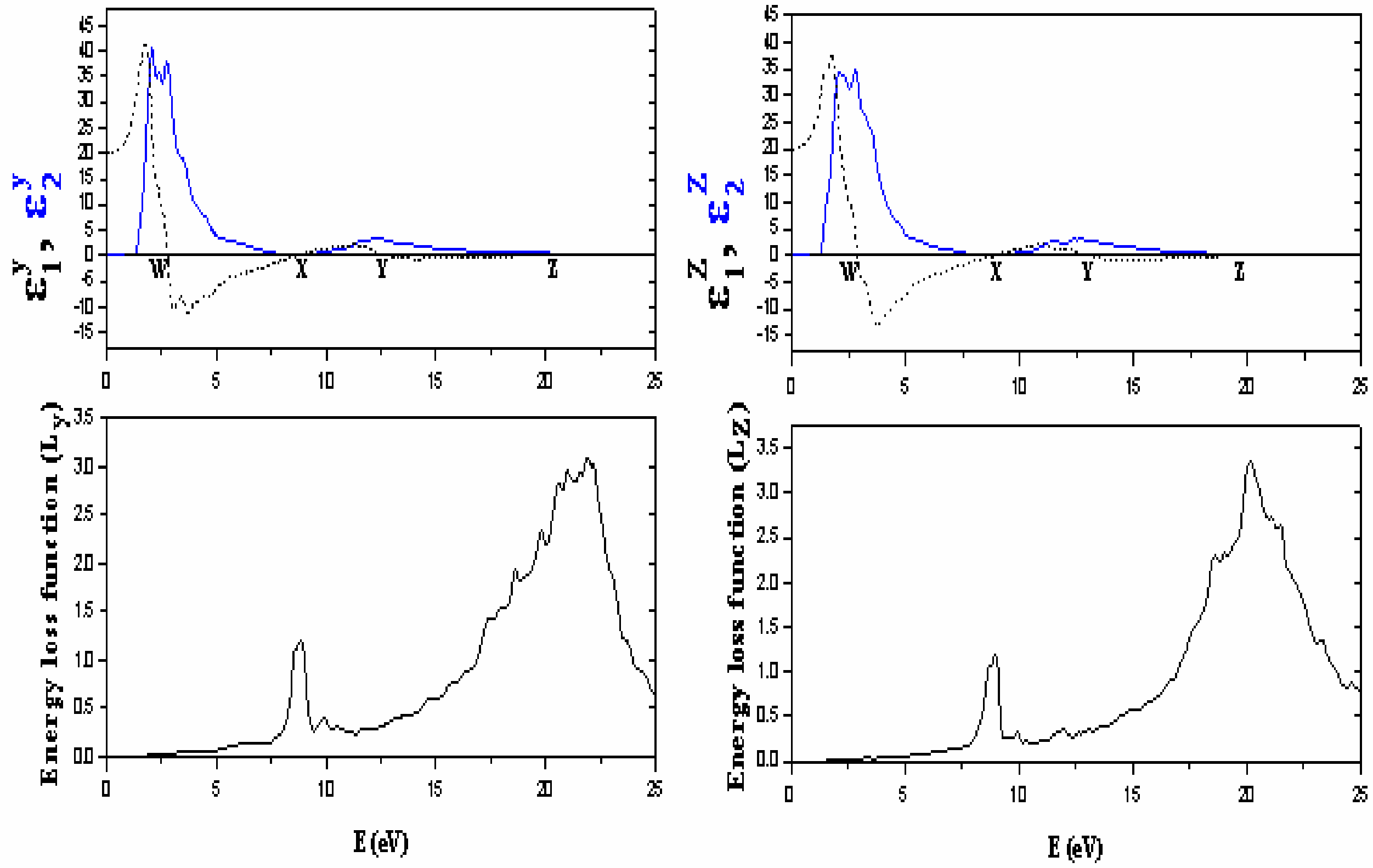

**Fig 3.** Energy spectra of dielectric function $\varepsilon = \varepsilon_1 - i\varepsilon_2$ and energy-loss function (L) along the y- and z-axes for $Sb_2Se_3$

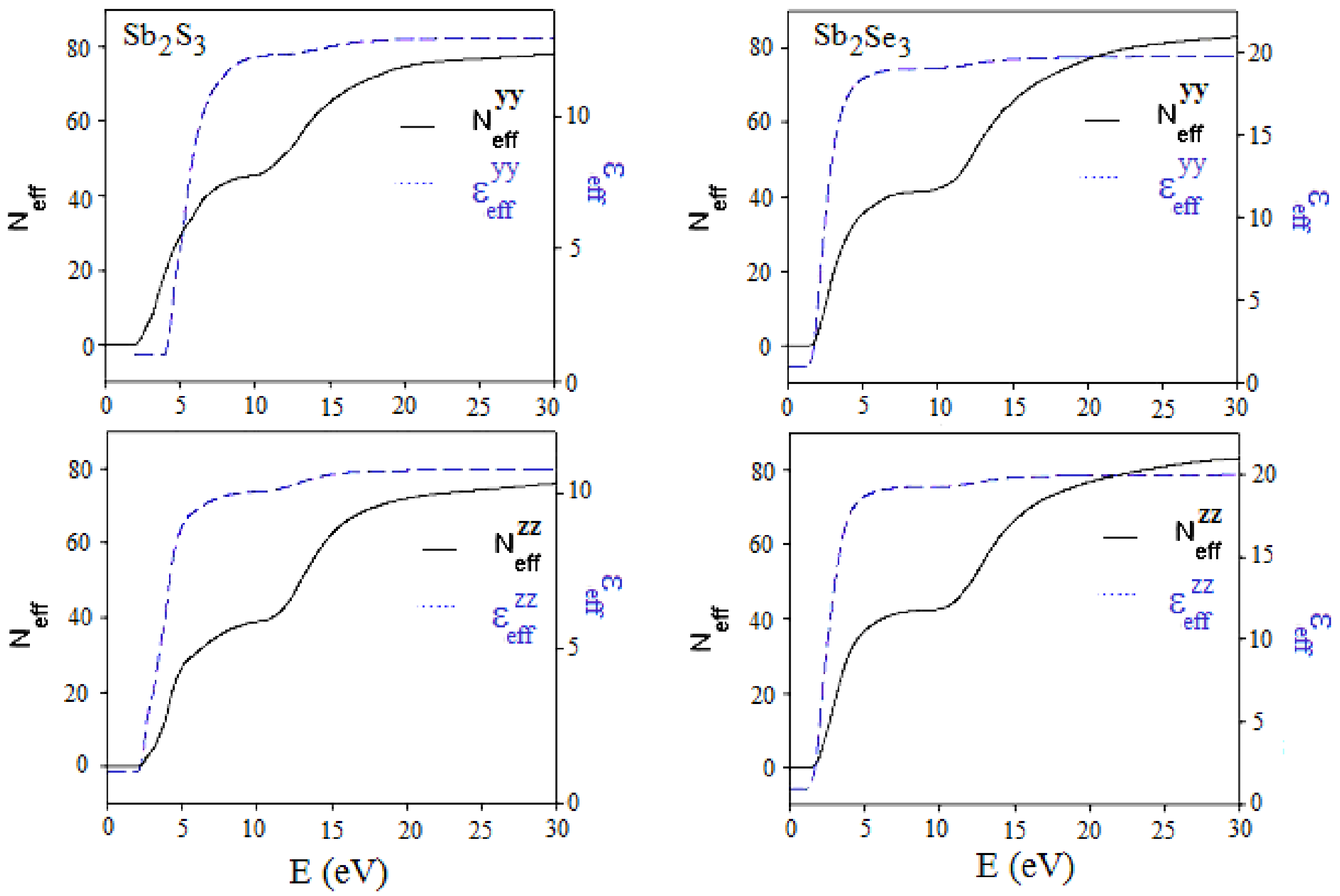

**Fig 4.** Energy spectra of $N_{eff}$ and $\varepsilon_{eff}$ along the y- and z- axes